\documentclass[prb,twocolumn]{revtex4-2} 
\usepackage{amsmath}  
\usepackage{amsfonts} 
\usepackage{graphicx} 
\usepackage{color}
\usepackage{cases} 
\usepackage{verbatim}
\usepackage{bm} 

\begin{document}

\title{Dynamics of certain Euler-Bernoulli rods and rings from a minimal coupling \\quantum isomorphism}

\author{T. A. Engstrom}
\email{tyler.engstrom@unco.edu} 
\affiliation{Department of Physics and Astronomy, University of Northern Colorado, Greeley, CO 80639, USA}

\date{\today}

\begin{abstract}
In some parameter and solution regimes, a minimally coupled nonrelativistic quantum particle in 1d is isomorphic to a much heavier, vibrating, very thin Euler-Bernoulli rod in 3d, with ratio of bending modulus to linear density $(\hbar/2m)^2$. For $m=m_e$, this quantity is comparable to that of a microtubule. Axial forces and torques applied to the rod play the role of scalar and vector potentials, respectively, and rod inextensibility plays the role of normalization. We show how an uncertainty principle $\Delta x\Delta p_x\gtrsim\hbar$ governs transverse deformations propagating down the inextensible, force and torque-free rod, and how orbital angular momentum quantized in units of $\hbar$ or $\hbar/2$ (depending on calculation method) emerges when the force and torque-free inextensible rod is formed into a ring. For torqued rings with large wavenumbers, a ``twist quantum'' appears that is somewhat analogous to the magnetic flux quantum. These and other results are obtained from a purely classical treatment of the rod, i.e., without quantizing any classical fields.
\end{abstract}

\maketitle

\section{Introduction}
Mathematical analogies between quantum mechanical systems and vibrating elastic systems have been made since the earliest days of quantum mechanics, when in 1926, Schr\"odinger observed that the ``square'' of what we now think of as the time-dependent Schr\"odinger equation resembles the equation of motion of a vibrating plate~\cite{schrodinger26, karam20}. Looking at this in reverse, the Euler-Bernoulli operator for a freely vibrating plate can be factored into two Schr\"odinger-like operators: $\partial_t^2+\Delta^2 = (i\partial_t+\Delta)(-i\partial_t+\Delta)$, a property useful in certain solution schemes~\cite{cordero11, orsingher11}, while it has similarly been shown that the F\"oppl-von K\'arm\'an operator describing large deflections of the plate can be factored into two nonlinear Schr\"odinger-like operators~\cite{hui-chuan87}. However, such analogies are fundamentally limited by the fact that the real, scalar transverse plate deflection or some derivative thereof stands in for a complex wavefunction. A closer continuum elastic analogy can be made to \emph{one-dimensional} quantum mechanics, and this involves a rod or elastica that deflects in two transverse dimensions. Here the components of the deflection vector (or even more powerfully, as we will see, the components of the transverse projection of the tangent vector) can play the roles of $\textrm{Re}[\Psi]$ and $\textrm{Im}[\Psi]$. While aspects of the 1d quantum - 3d rod analogy have appeared in the literature before, the focus has been on time-independent phenomena~\cite{odijk86, odijk98, hansen99, shi94, shi95, engstrom20}, specific forms of interaction relevant to semiflexible polymer systems~\cite{odijk86, odijk98, hansen99}, and nonlinear effects such as bend-twist coupling~\cite{shi94, shi95}, rather than on pushing the linear time-dependent analogy to the point of becoming an isomorphism, as we do here.

The value of such an isomorphism is, of course, in the potential to learn something about elastic systems from quantum systems, and vice versa, just as analogies between the band structure of electronic systems and vibrations of isostatic frames~\cite{kane14}, metamaterials~\cite{paulose15}, jammed packings~\cite{sussman16}, and continuous plates with patterened resonators~\cite{pal17} have recently been used to predict the existence of topologically protected edge modes in the latter four. As a more historic example, de Gennes showed that quantum fluctuations of a 1d Fermi gas can be mapped to thermal fluctuations of 2d polymeric systems (albeit flexible, not semiflexible polymers), where a ``no-crossing'' condition for polymer chains plays the role of Fermi statistics~\cite{degennes68}. While both the quantum and classical elastic systems considered here are comparatively simple, we nevertheless gain some surprising new insights. We find, for example, that de Broglie relations and uncertainty principles govern momentum-carrying transverse deformation pulses in periodic, inextensible, force and torque-free rods, and that vibrating inextensible rings may carry quantized orbital angular momentum reminiscent of spin. Additionally, a vibrating rod that experiences a spatially periodic body force, such as might be generated by internal interactions, can constitute a classical analog of a 1d Bloch electron (distinct from a phononic crystal in that the vibrations are flexural). The key to all this turns out to be the higher order and therefore virtually unused ``squared'' Schr\"odinger equation.

The rest of the paper is organized as follows. Section~\ref{section_formalism} develops the quantum side of the analogy and Section~\ref{section_rod} develops the elastic side, as generally as possible. Section~\ref{section_ring} concerns a specific elastic application, namely the ring or periodic rod mentioned above, for which the analogy lends considerable insight. We end in Section~\ref{section_discussion} with suggestions to use the analogy to explore new physics.

\section{Minimal coupling formalism} \label{section_formalism}
Consider a quantum particle with wavefunction $\Psi(\mathbf{r},t)=\Psi_1(\mathbf{r},t) + i\Psi_2(\mathbf{r},t)$, $\Psi_1$ and $\Psi_2$ being real, that satisfies the time-dependent Schr\"odinger equation  $i\hbar\partial_t\Psi = \hat H\Psi$. This problem is isomorphic to
\begin{equation}
\begin{pmatrix}
0 & -\hbar\partial_t \\
\hbar\partial_t & 0
\end{pmatrix}
\begin{pmatrix}
\Psi_1 \\
\Psi_2
\end{pmatrix}
 =  
\begin{pmatrix}
\textrm{Re}[\hat H] & -\textrm{Im}[\hat H] \\
\textrm{Im}[\hat H] & \textrm{Re}[\hat H]
\end{pmatrix}
\begin{pmatrix}
\Psi_1 \\
\Psi_2
\end{pmatrix}
,\label{rTDSE.0}
\end{equation}
written more concisely as
\begin{equation}
R\Big(\frac{\pi}{2}\Big)\hbar\partial_t\vec\Psi=\tilde H\vec\Psi, \label{rTDSE}
\end{equation}
where $R(\pi/2)$ is a $\pi/2$ rotation matrix, $\tilde H$ denotes the matrix of operators on the right hand side of Eq.~\eqref{rTDSE.0}, and $\vec\Psi(\mathbf{r},t)=(\Psi_1(\mathbf{r},t),\Psi_2(\mathbf{r},t))$ is a vector function in the now real, but $2\times$higher dimensional Hilbert space: $\mathbb{R}^2(\mathbb{R}^d)$ versus $\mathbb{C}(\mathbb{R}^d)$. (To avoid ambiguity, we use arrow notation and boldface for vectors in $\mathbb{R}^{2}$ and $\mathbb{R}^d$, respectively.) Transforming Eq.~\eqref{rTDSE} by operating on both sides with $R(\pi/2)\hbar\partial_t$ yields
\begin{equation}
-\hbar^2\partial_{tt}\vec{\Psi} = \tilde{H}^2\vec{\Psi} + R\Big(\frac{\pi}{2}\Big)\hbar[\partial_t, \tilde{H}]\vec{\Psi}. \label{SSE}
\end{equation}
When $[\partial_t, \tilde{H}]=0$, in other words when $\tilde{H}$ changes at most adiabatically slowly, Eq.~\eqref{SSE} is the real-valued counterpart of what is sometimes called the ``squared'' Schr\"odinger equation.

The Hamiltonian of interest in this work is the minimal coupling Hamiltonian
\begin{equation}
\hat H = \frac{1}{2m}\big(-i\hbar\boldsymbol{\nabla}-q\mathbf{A(r},t)\big)^2+V(\mathbf{r},t), \label{Hhat}
\end{equation}
corresponding to
\begin{equation}
\tilde H = -\frac{1}{2m} 
\begin{pmatrix}
\hbar\boldsymbol{\nabla} & q\mathbf{A} \\
-q\mathbf{A} & \hbar\boldsymbol{\nabla}
\end{pmatrix}^2
+
\begin{pmatrix}
V & 0 \\
0 & V
\end{pmatrix}. \label{Htilde}
\end{equation}
When $d=1$, $A=A_x$ is constant with respect to $x$ (e.g., Landau gauge), and $A$,$V$ have adiabatically slow or no time-dependence, Eq.~\eqref{SSE} with~\eqref{Htilde} can be written as the following set of coupled equations:

\begin{widetext}
\begin{subequations}
\label{rodEOMs}
\begin{eqnarray}
-\rho\ddot{\Psi}_1 &=& B\Psi_1'''' + M\Psi_2''' + (P\Psi_1')' + K\Psi_1 + C\Psi_2' + (C'/2)\Psi_2, \label{rodEOM1}\\
-\rho\ddot{\Psi}_2 &=& B\Psi_2'''' - M\Psi_1''' + (P\Psi_2')' + K\Psi_2 - C\Psi_1' - (C'/2)\Psi_1. \label{rodEOM2}
\end{eqnarray}
\end{subequations}
\end{widetext}
Dots and primes denote partial derivatives with respect to $t$ and $x$, respectively, while $B/\rho=(\hbar/2m)^2$, $M/\rho=\hbar qA/m^2$, $P/\rho=-V/m-3M^2/8\rho B$, $K=(P+M^2/4B)^2/4B + P''/2$, and $C=P(M/2B) + B(M/2B)^3$. 

\section{Dynamically and helically buckled rod} \label{section_rod}
Equations~\eqref{rodEOMs} come from an isomorphic rewriting of a problem in $d=1$ quantum mechanics, but in this Section, we attempt to reinterpret them as classical elastic equations of motion for an object living in $\mathbb{R}^{3}=\mathbb{R}^{2}\otimes\mathbb{R}^{d}$. Such a reinterpretation requires neglecting certain terms that have no clear physical meaning in the elastic context, yet these terms involve parameters that do have physical meaning in the elastic context. The conditions on the parameters (and solution regimes) under which these extra terms are negligible establish the limits of validity of the analogy.

With this philosophy in mind, we momentarily ignore the last two terms on the right hand sides of Eqs.~\eqref{rodEOMs}, and notice that what remains are the 3d equations of motion of an Euler-Bernoulli beam with circular cross section, hereafter, a rod. Table~\ref{tab:B/rho} suggests its scale. 
%
%
\begin{table}[h]
\caption{\label{tab:B/rho}Rods described by Eqs.~\eqref{rodEOMs} with $C=0$ are comparable in scale to semiflexible polymers, typically modeled as wormlike chains~\cite{broedersz14}. Examples of these are indicated with asterisks. Entries are in units of m$^4/$s$^2$.}
\begin{ruledtabular}
\begin{tabular}{l l l l l}
& & $(\hbar/2m)^2$ & $B/\rho$~\footnote{$B=\frac{\pi}{4}ER^4=k_BT\ell_P$, where $E$ is the rod's Young's modulus, $R$ is its radius, and $\ell_P$ is its persistence length in 3d. Thus $B/\rho$ scales as $R^2$ times the specific Young's modulus.} &\\ [1.0ex]
&proton & $9.9\times10^{-16}$ & &\\
&electron & $3.4\times10^{-9}$ & &\\
&DNA$^*$ & & $\sim10^{-13}$ &\\
&microtubule$^*$ & & $\sim10^{-10}$ &\\
&20 $\mu$m diameter glass fiber & & $\sim10^{-3}$ &\\
&16 ga. steel wire & & $\sim1$ &\\
\end{tabular}
\end{ruledtabular}
\end{table}
The rod undergoes transverse deflection from the $x$-axis proportional to $\vec{\Psi}$, has mass per unit length $\rho$ and bending modulus $B$, and sustains both an axial moment (torque) $M$ and an axial body force $P(x)$. ($P>0$ where the rod is in compression and $P<0$ where it is in tension.) Notice that $P$ accounts for twist-induced tension, suggesting the internal structure of the rod is a bundle of inextensible fibers~\cite{mertova18}. This picture is consistent with a single-valued bending modulus for the composite rod structure provided there is no cross-linking~\cite{broedersz14}. The rod also experiences a ``substrate'' force, e.g., from being embedded in an elastic gel, with force constant per unit length $K(x)$. The case of Eqs.~\eqref{rodEOMs} with $M=K=C=0$ and $P>0$ is known as the dynamical buckling equation~\cite{box20, kodio20, villermaux21}, yet the extra features described above are well-studied extensions of linear buckling~\cite{love44, landau86, silverberg12}, if not typically together and in a dynamical context. Counter-intuitively, even when a rod is everywhere in tension it can buckle given suitable boundary conditions~\cite{misseroni15}. The case of Eqs.~\eqref{rodEOMs} with $P=K=C=0$ describes flexural vibration of a rod with pre-twist~\cite{adair18}.

As $x\to0$, the second and fifth terms on the right hand sides of Eqs.~\eqref{rodEOMs} can be combined into $\pm(\mathcal{M}\Psi_{2,1}')''$ where $\mathcal{M}(x)=M+C(x)x^2/2$. So for part of the rod, the physical effect of $C$ is to generate a variable axial moment~\cite{silverberg12}. However, this coefficient doesn't appear to have a simple physical interpretation that is valid for the \emph{entire} rod, and so to make Eqs.~\eqref{rodEOMs} precisely analogous to an elastic system, we must neglect (and justify neglecting) the $C$ and $C'$ terms. To do this we introduce the rod length $L$ and note three different cases. 

\emph{Case 1:} Spatial derivatives of $P$ are negligible, and $ML/B$ and $PL^2/B$ are both $\ll1$ so that we may retain terms only to linear order in these quantities in the nondimensionalized equations of motion (lengths measured in units of $L$, time in units of $L^2\sqrt{\rho/B}$). On the quantum side, this case corresponds to the semiclassical approximation with weak potentials. On the elastic side, it corresponds to small terminal twist angle $\phi=(1+\nu)ML/B$ where $\nu$ is Poisson's ratio, and $P$ well below the Euler buckling threshold $\pi^2B/L^2$. Equations~\eqref{rodEOMs} reduce to 
\begin{subequations}
\label{rodEOMx.2}
\begin{eqnarray}
-\rho\ddot{\Psi}_1 &=& B\Psi_1'''' + M\Psi_2''' + P\Psi_1'' , \label{rodEOM1.2}\\
-\rho\ddot{\Psi}_2 &=& B\Psi_2'''' - M\Psi_1''' + P\Psi_2'' . \label{rodEOM2.2}
\end{eqnarray}
\end{subequations}
To linear order in $ML/B$, the twist-induced tension contribution to $P$ is not accounted for, and all elastic parameters become independent of one another. Incidentally, \emph{static} helically buckled rod configurations described by Eqs.~\eqref{rodEOMx.2} map onto the dynamics of a symmetric top that remains nearly upright and has its bottom point fixed, an instance of the Kirchhoff kinetic analogy~\cite{love44}.

\emph{Case 2:} For solutions composed entirely of short-wavelength ($\lambda\ll L$) Fourier modes, and for $P$ slowly varying on the scale of $\lambda$, the low-derivative-order terms on the right hand sides of Eqs.~\eqref{rodEOMs} can be negligible, even when their coefficients are not necessarily small. Specifically, for the $C$ terms to be negligible compared to the $M$ terms we require $(2\pi L/\lambda)^2\gg p/2+\phi^2/8$ and for the $K$ terms to be negligible compared to the $P$ terms we require $(2\pi L/\lambda)^2\gg p/4+\phi^2/8+\phi^4/64p$, where we have defined $p=PL^2/B$ and set $\nu=0$ for simplicity. This allows for modest values of twist $\phi$, for example four complete twists when $P$ is near the Euler threshold and $\lambda\ll L/4$, yet still corresponding to small torsional strain since the latter is equal to $\phi$ divided by the (extremely large) aspect ratio of the rod. Equations~\eqref{rodEOMx.2} also apply to this case.

\emph{Case 3:} $M=0$ (or removed via a gauge transformation) and $P(x)$ is an arbitrary function. Equations~\eqref{rodEOMs} uncouple and there is a special relationship between the elastic parameters: $K=P^2/4B+P''/2\approx P''/2$ when $P$ is small. In general the substrate term vanishes when $P$ is the solution of an Emden-Fowler equation.

\begin{figure}[t]
\centering
\includegraphics[width=0.48\textwidth]{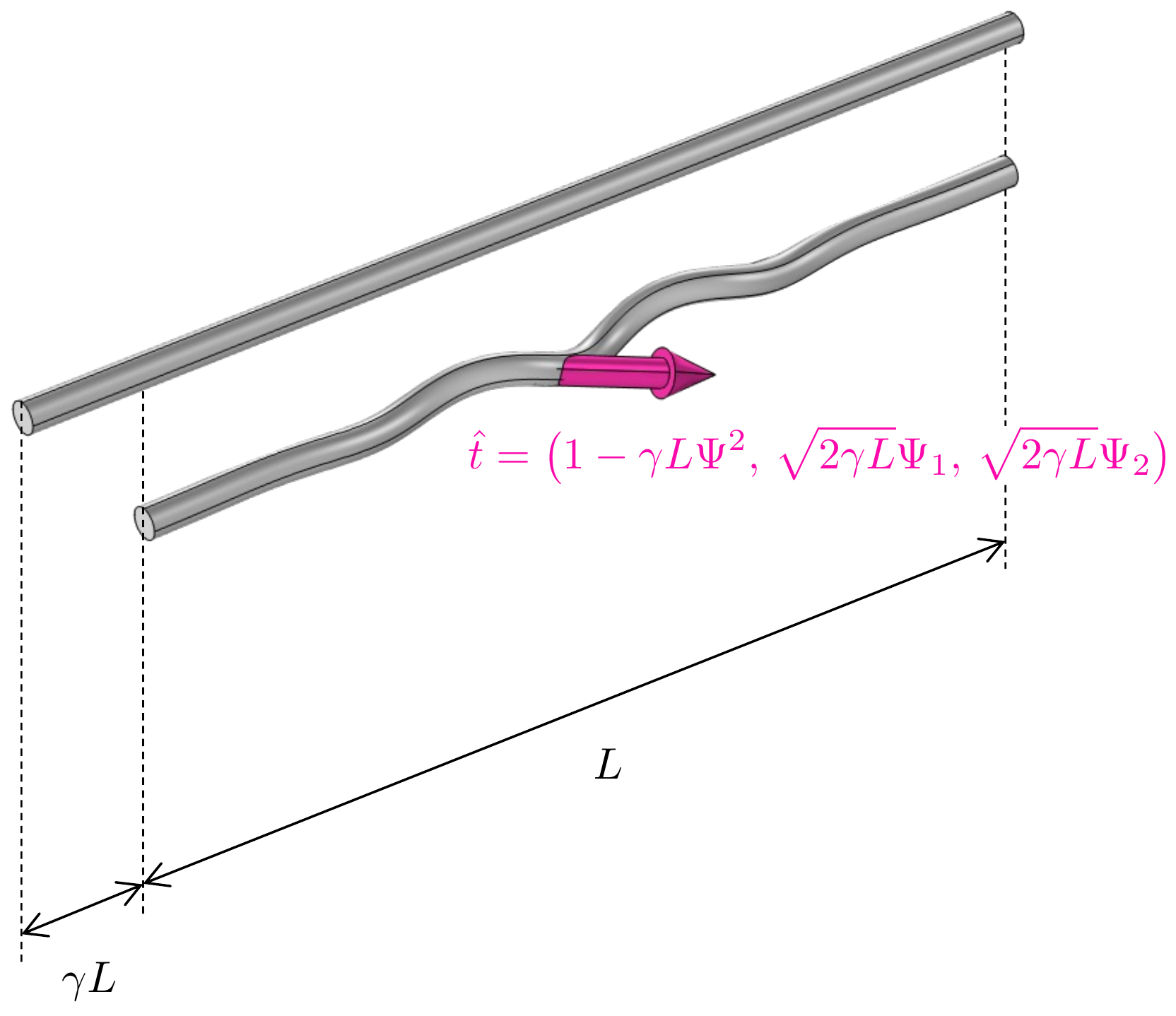}
\caption{\label{rod_schematic}Arbitrary small 3d deformation produced by subjecting a straight inextensible rod (oriented along $\hat x$) to  a relative compression $\gamma$ (shown greatly exaggerated). The vector function $\vec\Psi=\Psi_1\hat y + \Psi_2\hat z$ appearing in the expression for the rod's unit tangent vector $\hat t$ satisfies a ``squared'' Schr\"odinger equation (Eqs.\eqref{rodEOMx.2}) as well as a normalization-like constraint.}
\end{figure}

Here through Section~\ref{section_ring} we mostly focus on Cases 1 and 2, and at the end of Section~\ref{section_discussion} we return to Case 3. The essentially constant coefficients in Eqs.~\eqref{rodEOMx.2} mean that $\vec{\Psi}$ can now be interpreted as any scaled derivative of the rod's transverse deflection $\vec{u}(x,t)=u_1(x,t)\hat{y}+u_2(x,t)\hat{z}$. A particularly compelling choice is $\vec{\Psi}=\vec{u}'/\sqrt{2\gamma L}$, as this gives rise to a geometrical constraint $1=\int dx \vec{\Psi}\cdot\vec{\Psi}$ if the rod is inextensible, i.e., if the contour length of its centerline does not change. Here $L$ is redefined as the \emph{projected} (onto the $x$-axis) length of the rod, and $\gamma$ is its relative compression (not to be confused with axial strain, and not necessarily implying that $P>0$). Figure~\ref{rod_schematic} shows this geometry, and how $\vec\Psi$ becomes a scaled transverse projection of the rod's tangent vector.
%
%
The inextensibility constraint written above is the leading order term in an expansion and so is valid only when $\vec u'\cdot\vec u'\ll1$ everywhere, meaning $\gamma\ll1$. Inextensibility is a good assumption for semiflexible polymers --- it being an ingredient of the wormlike chain model~\cite{broedersz14} --- and is consistent with the internal structure of the rod inferred earlier, even though that structure is not accounted for at linear order in $ML/B$. So setting aside the broader analogy that can be made without specifying $\vec{\Psi}$ in favor of the deeper one that can be made by assigning $\vec{\Psi}=\vec{u}'/\sqrt{2\gamma L}$ to an inextensible rod, we proceed with the latter assignment except where otherwise noted. Thus, there exist ``normalized'' solutions of Eqs.~\eqref{rodEOMx.2} of the form
\begin{equation}
\vec{\Psi}(x,t) = \sum_n R(-\omega_nt)c_n\vec{\psi}_n(x), \label{gensol}
\end{equation}
where $(\tilde H/\hbar)\vec{\psi}_n=\omega_n\vec{\psi}_n$, $\delta_{nm}=\int dx\,\vec{\psi}_n\cdot\vec{\psi}_m$, $1=\sum_nc_n^2$, and in rod language,
\begin{equation}
\frac{\tilde H}{\hbar} = -\sqrt{\frac{B}{\rho}} 
\begin{pmatrix}
\partial_x^2+2\big(\tfrac{M}{4B}\big)^2+\tfrac{P}{2B} & 2\big(\tfrac{M}{4B}\big)\partial_x \\[6pt]
-2\big(\tfrac{M}{4B}\big)\partial_x & \partial_x^2+2\big(\tfrac{M}{4B}\big)^2+\tfrac{P}{2B}
\end{pmatrix}. \label{Htilde_v2}
\end{equation}
Note the identity $R(\alpha)=\exp[\alpha R(\pi/2)]$, Euler's formula for 2d rotations. The $\vec{\psi}_n$ are, in general, non-planar due to $M$, they rotate at angular frequencies $\omega_n$, and they correspond to stationary states. That is, if $\psi_n=\psi_{n,1}+i\psi_{n,2}$ is an eigenstate of $\hat H$ with eigenvalue $E_n$, then $\vec\psi_n=(\psi_{n,1}, \psi_{n,2})$ is an eigenstate of $\tilde H$ with the same eigenvalue. While the above described solution is not the only form of solution to Eqs.~\eqref{rodEOMx.2} --- more generally, the $\vec\psi_n$ satisfy the ``squared'' eigenvalue equation $(\tilde H/\hbar)^2\vec\psi_n=\omega_n^2\vec\psi_n$ --- we focus on this form since it is also a solution of the lower order Eq.~\eqref{rTDSE} with~\eqref{Htilde_v2} and therefore gives maximum contact with quantum mechanical methods and results.

The linear Eqs.~\eqref{rodEOMx.2} are agnostic as to whether $\rho$ is mass per unit length in the $x$-direction or mass per unit contour length (the local conversion factor being $1+\gamma L\Psi^2$), and therefore, we are free to choose the latter definition, which is the more realistic for an inextensible rod. A consequence is that purely transverse vibration of the inextensible rod is associated with a longitudinally propagating momentum. One way to see this is by combining the continuity equation $\partial_t[\rho(1+\gamma L \Psi^2)]+\partial_xg=0$ with Eqs.~\eqref{rTDSE} and \eqref{Htilde_v2} to obtain the momentum density
\begin{equation}
g(x,t) = -2\gamma L\sqrt{\rho B}\bigg[\vec{\Psi}\cdot
R\Big(\frac{\pi}{2}\Big)
\vec{\Psi}' + \frac{M}{4B}\Psi^2\bigg]. \label{pDens}
\end{equation}
Comparing to the quantum mechanical momentum density, i.e., $m$ times the probability current $-(\hbar/m)\{\textrm{Re}[\Psi^*i\Psi'] + (qA/\hbar)|\Psi|^2\}$, leads to the identification 
\begin{equation}
\hbar=2\gamma L\sqrt{\rho B}, \label{hbar}
\end{equation}
in terms of rod parameters. Putting this in the relationship $B/\rho=(\hbar/2m)^2$ further identifies $m=\gamma\rho L$. This result makes intuitive sense: the rod has mass $\rho L$ (to leading order), and so the pulse mass set in motion by a relative compression $\gamma$ is $\gamma\rho L$ (to leading order). In other words, the isomorphism connects a massive quantum particle with a much more massive (by a factor of $\gamma^{-1}$) elastic rod. 

For a field theoretic perspective on the rod's momentum density, we introduce the Lagrangian density 
\begin{equation}
\mathcal{L} = \frac{1}{2}\Big[\rho(\dot{\vec u})^2 - B(\vec u'')^2 - M\vec u'\cdot R\Big(\frac{\pi}{2}\Big)\vec u'' + P(\vec u')^2 \Big], \label{lagrangian}
\end{equation}
from which Eqs.~\eqref{rodEOMx.2} can be obtained. The last two terms represent energy stored by twisting and work done by $P$, and within Case 1, i.e., to linear order in $ML/B$, these terms are isomorphic to $-A_{\mu}j^{\mu}$ of classical electrodynamics. This follows from our previous results $qA_{\mu}=-\gamma L(P,\sqrt{\tfrac{\rho}{4B}}M,0,0)$ and $j^{\mu}/q\leftrightarrow (\Psi^2,g/\rho\gamma L,0,0)$. Calculating the relevant component of the energy-momentum tensor 
\begin{equation}
T_{\mu}^{\alpha} = \bigg[\frac{\partial\mathcal L}{\partial u_{i,\alpha}} - \partial_{\beta}\bigg(\frac{\partial\mathcal L}{\partial u_{i,\alpha\beta}}\bigg)\bigg]u_{i,\mu} + \frac{\partial\mathcal L}{\partial u_{i,\alpha\beta}}u_{i,\beta\mu} - \delta^{\alpha}_{\mu}\mathcal L,
\end{equation}
derived in Appendix~\ref{appendixA}, gives momentum density $g=-T_1^0=-\rho\dot{\vec u}\cdot\vec{u}'$, which matches Eq.~\eqref{pDens} when $M=P=0$. (To see this, use $\vec{u}$ as the dependent variable in Eq.~\eqref{rTDSE}, which is a valid procedure within Cases 1 and 2 as discussed above.) 

For yet a third perspective, consider the following geometric argument due to Rowland and Pask~\cite{rowland99}. Setting up a waveform in an initially straight inextensible rod requires that one end of the rod move by a distance $\gamma L=\frac{1}{2}\int_0^L dx(\vec u')^2$ toward the other, as shown in Figure~\ref{rod_schematic}. If the waveform has only one spectral component, then the solution has the property $\vec u'=-\dot{\vec u}/\dot x$, where $\dot x$ is the phase velocity (we will see a specific example in Section~\ref{section_ring}). Substituting this for one factor of $\vec u'$ in the integrand yields $\int dt\dot{(\gamma L)}=-\frac{1}{2}\int dt \dot{\vec u}\cdot\vec{u}'$, hence $g=-\frac{1}{2}\rho\dot{\vec u}\cdot\vec{u}'$, \emph{half} of that predicted by the other two methods when $M=P=0$. Rowland and Pask point out that this geometric argument gives the correct result for inextensible \emph{strings}, and in that linear dispersion context the waveform need not be restricted to one spectral component. 

\section{Slender inextensible ring \\or periodic rod} \label{section_ring}
To illustrate why the factor of $2$ discrepancy in $g$ poses an interesting question, we consider the $n^{th}$ helical eigenstate and corresponding eigenfrequency of Eq.~\eqref{Htilde_v2}:
\begin{subequations} \label{ring}
\begin{eqnarray}
&\vec\psi_n(x) = \frac{1}{\sqrt L}R(k_nx)\hat y, \label{e-state}\\
&\omega_n = \sqrt{\frac{B}{\rho}}\Big(\frac{2\pi}{L}\Big)^2\Big[\Big(n-\frac{\phi}{\phi_0}\Big)^2 - 3\Big(\frac{\phi}{\phi_0}\Big)^2 - \frac{p}{8\pi^2}\Big], 
\end{eqnarray}
\end{subequations}
where $k_n=2\pi n/L$, $M/4B=2\pi(\phi/\phi_0)/L$, and $\phi_0=8\pi(1+\nu)$. This describes a mode of vibration of a rod with periodic boundary conditions; alternatively, it describes a continuous ring of radius $L/2\pi$ if we neglect effects of the undeformed ring's curvature (compare the model in Ref.~\cite{box20}). Such neglect is justified when the ring is very slender ($R/L\to0$), and in any case, accounting for the curvature would introduce corrections to $P$ and $K$ that couple only to the in-plane component of $\vec\Psi$~\cite{landau86, kodio20} --- interesting, but incompatible with the isomorphism here. The ring deflection corresponding to Eqs.~\eqref{ring} is, up to a rigid translation of the guiding center,
\begin{equation}
\vec u_n(x,t)=-\frac{\sqrt{2\gamma}}{k_n}R(k_nx-\omega_nt)\hat z, \label{ring_shape}
\end{equation}
and one can readily verify that $\vec u_n'=-(k_n/\omega_n)\dot{\vec u}_n$ as required by the geometric argument.

Letting the plane of the ring be perpendicular to the $z$-axis, $(L/2\pi)\int_0^L dx g$ yields orbital angular momentum
\begin{widetext}
\begin{equation} \label{Lz}
L_z=
\begin{cases}
\hbar\big(n-\frac{\phi}{\phi_0}\big), & g\textrm{ from continuity equation,}\\[6pt]
\hbar\big(n-2\frac{\phi}{\phi_0}-\frac{1}{n}\big[2\big(\frac{\phi}{\phi_0}\big)^2 + \frac{p}{8\pi^2}\big]\big), & g\textrm{ from energy-momentum tensor,}\\[6pt]
\hbar\big(\frac{n}{2}-\frac{\phi}{\phi_0}-\frac{1}{n}\big[\big(\frac{\phi}{\phi_0}\big)^2 + \frac{p}{16\pi^2}\big]\big), & g\textrm{ from geometric argument.}
\end{cases}
\end{equation}
\end{widetext}
The first two methods predict $L_z$ is quantized in units of $\hbar$ when $M=P=0$, while the third method intriguingly predicts $L_z$ is quantized in units of $\hbar/2$ when $M=P=0$. Of course, when we say ``units of $\hbar$'' we are assuming the factor of length change $\gamma L$ in $\hbar$ is independent of $n$, which is not intuitive for elastic rings. And yet the isomorphism is describing a particular elastic ring for which $\gamma L$ is indeed independent of $n$. Next we consider nonzero $M$ and $P$ within a sub-case of Case 2 corresponding to $4\pi^2 L/\lambda \gg p/2+\phi^2/8$. Within this sub-case the terms $\sim1/n$ in Eq.~\eqref{Lz} are negligible, and the $L_z$ spectrum becomes periodic in $\phi$ with period $\phi_0$ or $\phi_0/2$, depending on method. Thus $\phi_0$ (or $\phi_0/2$) plays a role comparable to that of the flux quantum for a conducting loop in a magnetic field~\cite{altland06,byers61}, just not in a thermal equilibrium sense since here the total energy (found by changing the signs of the last three terms in Eq.~\eqref{lagrangian} and integrating over $x$) is not also periodic in $\phi$. 

Letting Eqs.~\eqref{ring} describe the periodic straight rod and setting $M=P=0$ yields a set of de Broglie relations: $p_x=\int dxg=\hbar k_n$ (using one of the first two methods for $g$) and $E=\hbar\omega_n$ (equal parts kinetic and potential energy and also applicable to the ring). There is also orbital angular momentum $L_x = \int dx \rho u_n^2\omega_n = \hbar$, which makes the rod reminiscent of an OAM beam~\cite{allen92}. Now suppose the rod is sufficiently long that we can take the $k_n$ to be continuous. The general $t=0$ state of the rod is given by
\begin{equation}
\vec{\Psi}(x,0)=\frac{1}{\sqrt{2\pi}}\int dk\,R(kx) \vec{\Omega}(k),
\end{equation}
where $\vec{\Omega}(k)$ is the continuum generalization of $c_n\hat{y}$ in Eqs.~\eqref{gensol} and \eqref{e-state}. The relationship between $\vec{\Psi}(x,0)$ and $\vec{\Omega}(k)$ is isomorphic to that of a Fourier transform pair, as shown in Appendix~\ref{appendixB}. Thus if $\vec{\Omega}(k)$ is a peaked vector function with characteristic peak width $\Delta k$, then $\vec{\Psi}(x,0)$ describes a transverse deformation with characteristic length $\Delta x$ (as in the lower image of Figure~\ref{rod_schematic}), and by a property of Fourier transforms, $\Delta x \Delta k \gtrsim1$. This inequality is generic and unremarkable, but when combined with the first de Broglie relation for the rod, it becomes the rather more remarkable Heisenberg uncertainty principle
\begin{equation}
\Delta x\Delta p_x \gtrsim \hbar. \label{uncertainty}
\end{equation}
An energy-time uncertainty principle follows from the other de Broglie relation: a transverse deformation pulse of spatial extent $\Delta x$ moving at the group velocity $\partial\omega/\partial k\sim\Delta\omega/\Delta k$ takes time $\Delta t \sim \Delta x\Delta k/\Delta\omega \gtrsim 1/\Delta\omega$ to pass a particular point, hence
\begin{equation}
\Delta t\Delta E \gtrsim \hbar. \label{uncertainty2}
\end{equation}

\section{Discussion} \label{section_discussion}
Conventional wisdom has it that Eq.~\eqref{uncertainty} is fundamentally quantum mechanical, foreign to classical mechanics. While this may be true for classical mechanics of \emph{particles}, it is evidently not so for classical mechanics of elastic \emph{continua}. One might argue that $\hbar$ would not appear in any purely classical theory, but we again point to Table~\ref{tab:B/rho} which demonstrates that a rod with ratio of bending modulus to linear density equal to $(\hbar/2m_e)^2$ is comparable to a microtubule or in fact something a bit heavier --- a classical object indeed. And recall that the right hand side of Eq.~\eqref{uncertainty} can be written purely in terms of classical rod parameters, per Eq.~\eqref{hbar}.

It is somewhat unsatisfying that three different methods give three different results for $g$, and thus the inextensible ring's $L_z$, but a few comments are in order. First, the geometric argument makes use of a $\gamma=0$ reference state that is distinct from that of other two methods, and also appears to have no counterpart in the quantum problem. (A complex wavefunction that is ``stretched out'' beyond the extent of its real-space domain is an unphysical concept.) Second, the previously mentioned equipartition of energy when $M=P=0$ suggests that half of the $g$ obtained from the first two methods may actually be  a pseudo-momentum density associated with potential energy transport. This has also been suggested for inextensible strings~\cite{rowland99}. Third, recall that quantum mechanics itself suffers from method dependence: half-integer angular momenta defy an apparently rigorous separation of variables method but yield to an algebraic method. 

Indeed, the possibility of the inextensible ring's $L_z$ being quantized in units of $\hbar/2$ hints at the notion of the ring as a particle-like entity. We have seen that this ``particle'' would have mass greater by a factor of $\gamma^{-1}$ than the mass $m$ of some corresponding particle. If $m=m_e$, which again suggests taking $B$ and $\rho$ values from a microtubule, then we may crudely estimate $\gamma L=\hbar/2\sqrt{\rho B}\sim10^{-17}$~m, and crudely estimate the ring's mass to radius ratio as $2\pi m_e/\gamma L\sim10^{-12}$~kg/m.  The slenderness requirement ($L\gg$ microtubule radius) leads to an estimate $\gamma\ll10^{-9}$; for $\gamma$ this small, the Euler-Bernoulli description and the form of the inextensibility constraint we have used are essentially exact. Attempting to further constrain $\gamma$ and $L$ would be an interesting direction for future work.

Another future direction concerns the analog of wavefunction collapse. Conventionally in dynamical buckling (where $M=K=0$ and $P$ is a positive constant), one assumes that a perturbation with wave number $k$ grows at rate $\sigma(k)$. This leads to a prediction that the fastest-growing, hence, most probable mode is the one with $k=\sqrt{P/2B}$, a result that is in good agreement with experiments on rings confined to buckle in plane~\cite{box20}. (As the authors point out, this agreement can be taken as evidence that the ring curvature is negligible for slender rings, as we have assumed in Section~\ref{section_ring}.) The natural question to ask in the present context, however, is not which mode is selected out of an unbuckled state, but which mode is selected out of a superposition state? Because Eq.~\eqref{uncertainty} is linked to the statistical interpretation in quantum mechanics, its presence here also hints at a statistical interpretation, i.e., that the $c_n^2$ may be probabilities of the outcomes of measurements, such as of the straight rod's $L_x = \int dx \rho u^2\omega$.

Finally, let us return to Case 3, where $\vec\Psi\sim$ rod deflection and $P(x)$ is an arbitrary function. A convenient property of the isomorphism is that if $PL^2/B$ is a weak sinusoidal function, so is $(KL^4/B)\approx (PL^2/2B)''$. This suggests \emph{engineering} out-of-phase sinusoidal variations in the axial body force and in the substrate stiffness. One way to achieve the former could be by manipulating intrachain interactions in a stiff polyelectrolyte --- already implicated in buckling~\cite{hansen99} --- by creating an alternating pattern of charged and neutral monomer blocks~\cite{lytle19}. If such a polymer were made to vibrate in a non-dissipative environment of similar polymers, all parallel to and in registry with the first, the effective $K$ from that environment should oscillate out of phase with $P$ and be tunable in strength through the environment density. Eigenstates of Eq.~\eqref{Htilde_v2} would then be expected to have the Bloch form
\begin{equation}
\vec\psi_{n,k}(x)\sim R(kx)\vec w_{n,k}(x),
\end{equation}
where $n$ is a band index and the function $\vec w$ has the periodicity of the monomer pattern. Assuming such a system could be realized experimentally, one might then imagine adiabatically varying the boundary conditions --- certain of which are equivalent to twist via a gauge transformation --- to search for more exotic physics such as the Zak phase~\cite{zak89}, but in a classical setting.

\appendix
\section{energy-momentum tensor}\label{appendixA}
The energy-momentum tensor, a.k.a. the stress-energy tensor, is usually derived assuming the Lagrangian density depends on field variable derivatives only up to first order. Here we derive it (in the notation of Ref.~\cite{jose98}) for a Lagrangian density such as Eq.~\eqref{lagrangian} that depends on field variable $u(x)$ derivatives up to second order: $\mathcal L=\mathcal L(u, \nabla u, \nabla\nabla u, x)$. The Euler-Lagrange equations are
\begin{equation}
\frac{\partial\mathcal L}{\partial u_i} 
- \partial_{\alpha}\frac{\partial\mathcal L}{\partial u_{i,\alpha}} 
+ \partial_{\alpha\beta}\frac{\partial\mathcal L}{\partial u_{i,\alpha\beta}} = 0. \label{EL}
\end{equation}
Under an $\epsilon$-family of transformations of the field variables, and writing $\delta\equiv d/d\epsilon$ at $\epsilon=0$,
\begin{equation}
\delta\mathcal L = \frac{\partial\mathcal L}{\partial u_i}\delta u_i + \frac{\partial\mathcal L}{\partial u_{i,\alpha}}\delta u_{i,\alpha} + \frac{\partial\mathcal L}{\partial u_{i,\alpha\beta}}\delta u_{i,\alpha\beta}.
\end{equation}
Combining the previous two equations yields
\begin{equation}
\delta\mathcal L = \partial_{\alpha}\bigg[\bigg{\{}\frac{\partial\mathcal L}{\partial u_{i,\alpha}} - \partial_{\beta}\bigg(\frac{\partial\mathcal L}{\partial u_{i,\alpha\beta}}\bigg)\bigg{\}}\delta u_i + \frac{\partial\mathcal L}{\partial u_{i,\alpha\beta}}\delta u_{i,\beta} \bigg].
\end{equation}
Now if $\epsilon$ parameterizes a displacement of the origin by an arbitrary four-vector $h^{\mu}$, i.e., $u_i(x^{\mu})\to u_i(x^{\mu}+\epsilon h^{\mu})$, then $\delta u_i=h^{\mu}\partial_{\mu}u_i$ and $\delta\mathcal L=h^{\mu}\partial_{\mu}\mathcal L=\partial_{\alpha}(h^{\mu}\delta^{\alpha}_{\mu}\mathcal L)$. Hence
\begin{equation}
0 = \partial_{\alpha}T_{\mu}^{\alpha},
\end{equation}
where
\begin{equation}
T_{\mu}^{\alpha} = \bigg[\frac{\partial\mathcal L}{\partial u_{i,\alpha}} - \partial_{\beta}\bigg(\frac{\partial\mathcal L}{\partial u_{i,\alpha\beta}}\bigg)\bigg]u_{i,\mu} + \frac{\partial\mathcal L}{\partial u_{i,\alpha\beta}}u_{i,\beta\mu} - \delta^{\alpha}_{\mu}\mathcal L.
\end{equation}
\section{Fourier transform analog}\label{appendixB}
Due to the existence of an Euler's formula for 2d rotations: $\exp[\alpha R(\pi/2)]=I\cos\alpha+R(\pi/2)\sin\alpha=R(\alpha)$, we can write down analogs of the Fourier transform for vector functions in $\mathbb{R}^{2}$. These are
\begin{subequations}
\begin{eqnarray}
\vec{\xi}(x) &=& \frac{1}{\sqrt{2\pi}}\int dk\,R(kx)\vec{\xi}(k),\\
\vec{\xi}(k) &=& \frac{1}{\sqrt{2\pi}}\int dx\,R(-kx)\vec{\xi}(x),
\end{eqnarray}
\end{subequations}
with Dirac delta functions given by
\begin{subequations}
\begin{eqnarray}
I\delta(x) &=& \frac{1}{2\pi}\int dk\,R(kx),\\
I\delta(k) &=& \frac{1}{2\pi}\int dx\,R(kx).
\end{eqnarray}
\end{subequations}

\begin{acknowledgments}
The author thanks Ted Allen and Jan Chaloupka for useful comments on an early version of this manuscript, and David Rowland for an enlightening discussion about momentum density of strings and rods.
\end{acknowledgments}

\end{document}